\documentclass[prb,twocolumn,groupedaddress]{revtex4} % FOR JCP SUBMISSION

\usepackage{graphicx, bm, amsfonts, amssymb, amsmath, appendix}
\usepackage[colorlinks=true,citecolor=blue,linkcolor=blue]{hyperref}

\graphicspath{{figures/}}

\newcommand{\T}{\mathrm{T}}                                % T used in matrix transpose

\begin{document}

\title{Optimal estimators and asymptotic variances for nonequilibrium path-ensemble averages}

\author{David D. L. Minh}
 \email[Electronic Address: ]{daveminh@gmail.com}
 \affiliation{Laboratory of Chemical Physics, NIDDK, National Institutes of Health, Bethesda, Maryland 20892, USA}
\author{John D. Chodera}
\email[Electronic Address: ]{jchodera@berkeley.edu}
\affiliation{Research Fellow, California Institute of Quantitative Biomedical Research (QB3), University of California, Berkeley, 260J Stanley Hall, Berkeley, California 94720, USA}

\date{\today}

%%%%%%%%%%%%%%%%%%%%%%%%%%%%%%%%%%%%%%%%%%%%%%%%%%%%%%%%%%%%%%%%%%%%%%
% ABSTRACT
%%%%%%%%%%%%%%%%%%%%%%%%%%%%%%%%%%%%%%%%%%%%%%%%%%%%%%%%%%%%%%%%%%%%%%

\begin{abstract}
Existing optimal estimators of nonequilibrium path-ensemble averages are shown to fall within the framework of extended bridge sampling.  Using this framework, we derive a general minimal-variance estimator that can combine nonequilibrium trajectory data sampled from multiple path-ensembles to estimate arbitrary functions of nonequilibrium expectations.  The framework is also applied to obtaining asymptotic variance estimates, which are a useful measure of statistical uncertainty.  In particular, we develop asymptotic variance estimates pertaining to Jarzynski's equality for free energies and the Hummer-Szabo expressions for the potential of mean force, calculated from uni- or bidirectional path samples.  Lastly, they are demonstrated on a model single-molecule pulling experiment.  In these simulations, the asymptotic variance expression is found to accurately characterize the confidence intervals around estimators when the bias is small.  Hence, it does not work well for unidirectional estimates with large bias, but for this model it largely reflects the true error in a bidirectional estimator derived by Minh and Adib.
\end{abstract}

\maketitle

%%%%%%%%%%%%%%%%%%%%%%%%%%%%%%%%%%%%%%%%%%%%%%%%%%%%%%%%%%%%%%%%%%%%%%
% INTRODUCTION
%%%%%%%%%%%%%%%%%%%%%%%%%%%%%%%%%%%%%%%%%%%%%%%%%%%%%%%%%%%%%%%%%%%%%%

\section{Introduction}

Path-ensemble averages play a central role in nonequilibrium statistical mechanics, akin to the role of configurational ensemble averages in equilibrium statistical mechanics.  
Expectations of various functionals over processes where a system is driven out of equilibrium by a time-dependent external potential have been shown to be related to equilibrium properties, including free energy differences \cite{Jarzynski1997a, Jarzynski1997b} and thermodynamic expectations. \cite{Crooks2000, Neal2001}
The latter relationship, between equilibrium and nonequilibrium expectations, has been applied to several specific cases, such as: 
the potential of mean force (PMF) along the pulling coordinate \cite{Hummer2001a, Hummer2005, Minh2006} 
(or other observed coordinates \cite{Minh2007}) in single-molecule pulling experiments; 
RNA folding free energies as a function of a control parameter; \cite{Junier2009}
the root mean square deviation from a reference structure; \cite{Lyman2007}
the potential energy distribution \cite{Lyman2007} and average; \cite{Nummela2008} 
and the thermodynamic length. \cite{Feng2009}

Compared to equilibrium sampling, nonequilibrium processes may be advantageous for traversing energetic barriers and accessing larger regions of phase space per unit time.
This is useful, for example, in reducing the effects of experimental apparatus drift or increasing the sampling of barrier-crossing events.
Thus, there has been interest in calculating equilibrium properties from nonequilibrium trajectories collected in simulations or laboratory experiments.
Indeed, single-molecule pulling data has been used to experimentally verify relationships between equilibrium and nonequilibrium quantities. \cite{Liphardt2002, Collin2005}

While many estimators for free energy differences \cite{Bennett1976, Crooks2000, Maragakis2006, Maragakis2008} and equilibrium ensemble averages can be constructed from nonequilibrium relationships, they will differ in the efficiency with which they utilize finite data sets, leading to varying amounts of statistical bias and uncertainty.
Characterization of this bias and uncertainty is helpful for comparing the quality of different estimators \cite{Shirts2005} and assessing the accuracy of a particular estimate.
The statistical uncertainty of an estimator is usually quantified by its variance in the asymptotic, or large sample, limit, where estimates from independent repetitions of the experiment often approach a normal distribution about the true value due to the central limit theorem.
It is an important goal to find an optimal estimator which minimizes this asymptotic variance.

Although numerical estimates of the asymptotic variance may be provided by bootstrapping (e.g.~Ref.~\cite{Calderon2009}), closed-form expressions can provide computational advantages in the computation of confidence intervals, allow comparison of asymptotic efficiency, \cite{Tan2004, Shirts2005} and facilitate the design of adaptive sampling strategies to target data collection in a manner that most rapidly reduces statistical error. \cite{Singhal2005,Singhal2007,Hahn2009}
In the asymptotic limit, the statistical error in functions of the estimated parameters can be estimated by propagating this variance estimate via a first-order Taylor series expansion.
While this procedure is relatively straightforward for simple estimators, it can be difficult for estimators that involve arbitrary functions (e.g. nonlinear or implicit equations) of nonequilibrium path-ensemble averages.

Fortunately, the extended bridge sampling (EBS) estimators, \cite{Vardi1985, Gill1988, Kong2003, Tan2004} a class of equations for estimating the ratios of normalizing constants, are known to have both minimal-variance forms and associated asymptotic variance expressions. 
Recently, Shirts and Chodera \cite{Shirts2008} applied the EBS formalism to generalize the Bennett acceptance ratio, \cite{Bennett1976} producing an optimal estimator combining data from multiple equilibrium states to compute free energy differences, thermodynamic expectations, and their associated uncertainties.  
Here, we apply the EBS formalism to estimators utilizing nonequilibrium trajectories.  
We first construct a general minimal-variance path-average estimator that can use samples collected from multiple nonequilibrium path-ensembles.
We then show that some existing path-average estimators using uni- and bidirectional data are special cases of this general estimator, proving their optimality.
This also allows us to develop asymptotic variance expressions for estimators based on Jarzynski's equality \cite{Jarzynski1997a, Jarzynski1997b} and the Hummer-Szabo expressions for the PMF. \cite{Hummer2001a, Hummer2005, Minh2008prl}
We then demonstrate them on simulation data from a simple one-dimensional system and comment on their applicability.

%%%%%%%%%%%%%%%%%%%%%%%%%%%%%%%%%%%%%%%%%%%%%%%%%%%%%%%%%%%%%%%%%%%%%%
% EXTENDED BRIDGE SAMPLING
%%%%%%%%%%%%%%%%%%%%%%%%%%%%%%%%%%%%%%%%%%%%%%%%%%%%%%%%%%%%%%%%%%%%%%

\section{Extended Bridge Sampling}

Suppose that we sample $N_i$ paths (trajectories) from each of $K$ path-ensembles indexed by $i = 1,2,...,K$.  
The path-ensemble average of an arbitrary functional $\mathcal F[X]$ in path-ensemble $i$ is defined by
\begin{equation}
\left< \mathcal F \right>_i \equiv \int dX \, \mathcal F[X] \, \rho_i[X],
\label{eq:path_average}
\end{equation}
where $\rho_i[X]$ is a probability density over trajectories,
\begin{eqnarray}
\rho_i[X] = c_i^{-1} q_i[X] \:\: ; \:\: c_i = \int dX \, q_i[X],
\label{eq:path_density}
\end{eqnarray}
with unnormalized density $q_i[X] > 0$ and the normalization constant $c_i$ (a path partition function).
The above integrals, in which $dX$ is an infinitesimal path element, are taken over all possible paths, $X$.  
Extended bridge sampling estimators provide a way of estimating ratios of normalization constants $c_i/c_j$, which will prove useful in estimating free energies and thermodynamic expectations.

To construct these estimators, we first note the importance sampling identity,
\begin{eqnarray}
c_i \left< \alpha_{ij} \, q_j \right>_i 
& = & \left[ \int dX \, q_i[X] \right] \frac{ \int dX \, \alpha_{ij}[X] \, q_i[X] \, q_j[X] }{ \int dX \, q_i[X] } \nonumber \\
& = & \left[ \int dX \, q_j[X] \right] \frac{ \int dX \, \alpha_{ij}[X] \, q_i[X] \, q_j[X] }{ \int dX \, q_j[X] } \nonumber \\
& = & c_j \left< \alpha_{ij} \, q_i \right>_j,
\label{equation:importance-sampling-identity}
\end{eqnarray}
where $j$ is another path-ensemble index, $\alpha_{ij}[X]$ is an arbitrary functional of $X$, and all normalization constants are nonzero.

Summing over the index $j$ in Eq.\ \ref{equation:importance-sampling-identity} and using the sample mean, $N_i^{-1} \sum_{n=1}^{N_i} \mathcal F[X_{in}]$, as an estimator for $\left< \mathcal F \right>_i$, we obtain a set of $K$ estimating equations,
\begin{eqnarray}
\sum_{j=1}^{K} \frac{\hat{c}_i}{N_i} \sum_{n=1}^{N_i} \alpha_{ij}[X_{in}] \, q_j[X_{in}] = 
\sum_{j=1}^{K} \frac{\hat{c}_j}{N_i} \sum_{n=1}^{N_j} \alpha_{ij}[X_{jn}] \, q_i[X_{jn}],
\label{eq:ext_bridge}
\end{eqnarray}
whose solutions yield estimates $\hat{c}_i$ for the normalization constants $c_i$, up to an irrelevant scalar multiple. 
Each path, $X_{in}$, is indexed by the ensemble $i$ from which it is sampled, and the sample number $n = 1,2,...,N_i$.  
This coupled set of nonlinear equations defines a \emph{family} of estimators parameterized by the choice of $\alpha_{ij}[X]$, all of which are asymptotically consistent, but whose statistical efficiencies will vary. \cite{Tan2004}

With the choice,
\begin{equation}
\alpha_{ij}[X] = \frac{N_j \hat{c}_j^{-1} }{ \sum\limits_{k=1}^K N_k \, \hat{c}_k^{-1} \, q_k[X]},
\end{equation}
Eq.\ \ref{eq:ext_bridge} simplifies to the optimal EBS estimator,
\begin{equation}
\hat{c}_i = \sum_{j=1}^K \sum_{n=1}^{N_j} 
\left[ \sum_{k=1}^K \frac{N_k}{\hat{c}_k} \frac{q_k[X_{jn}]}{q_i[X_{jn}]} \right]^{-1}. \label{eq:opt_bridge}
\end{equation}
This choice for $\alpha_{ij}[X]$ is optimal in that the asymptotic variance of the ratios $\hat{c}_i / \hat{c}_j$ is minimal. \cite{Tan2004,Shirts2008}
These equations may be solved by any appropriate algorithm, including a number of efficient and stable methods suggested by Shirts and Chodera. \cite{Shirts2008}

The asymptotic covariance of Eq.\ \ref{eq:opt_bridge} is estimated by,
\begin{equation}
\hat{\bm{\Theta}} = \bm{M}^\T(\bm{I}_N - \bm{MNM}^\T)^+ \bm{M}
\label{eq:Theta}
\end{equation}
where the elements of $\bm{\Theta}$ are the covariances of the logarithms of the estimated normalization constants, 
$\Theta_{ij} = \mathrm{cov}\,(\hat{\gamma}_i, \hat{\gamma}_j)$, 
and $\hat{\gamma}_i = \ln \hat{c}_i$.\cite{Kong2003}
The superscript $(...)^+$ denotes an appropriate generalized inverse, such as the Moore-Penrose pseudoinverse, 
$\bm{I}_N$ is the $N \times N$ identity matrix 
(where $N=\sum_{i=1}^K N_i$ is the total number of samples),
$\bm{N} = \mathrm{diag} \, (N_1,N_2,...,N_K)$ is the diagonal matrix of sample sizes, and $\bm{M}$ is the $N \times K$ weight matrix with elements,
\begin{equation}
M_{ni} = \hat{c}_i^{-1} \frac{ q_i[X_n] }{ \sum\limits_{k=1}^K N_k \, \hat{c}_k^{-1} \, q_k[X_n] }.
\label{eq:weight_elements}
\end{equation}
In this matrix, the distribution from which samples are drawn from is irrelevant, and $X$ is only indexed by $n = 1,\ldots,N$.  We note that the sum over each column, $\sum_{n=1}^{N} M_{ni}$, is one.

For arbitrary functions of the logarithms of the normalization constants, $\phi(\hat{\gamma}_1,...,\hat{\gamma}_K)$ and $\psi(\hat{\gamma}_1,...,\hat{\gamma}_K)$, the asymptotic covariance $\mathrm{cov}(\hat{\phi},\hat{\psi})$ can be estimated from $\hat{\bm{\Theta}}$ according to,
\begin{equation}
\mathrm{cov} (\hat{\phi},\hat{\psi}) \approx
\sum_{i,j=1}^K \frac{ \partial \phi }{ \partial \hat{\gamma}_i } \hat{\Theta}_{ij} \frac{ \partial \psi }{ \partial \hat{\gamma}_j },
\label{eq:cov}
\end{equation}
through first-order Taylor series expansion of $\phi$ and $\psi$. 

%%%%%%%%%%%%%%%%%%%%%%%%%%%%%%%%%%%%%%%%%%%%%%%%%%%%%%%%%%%%%%%%%%%%%%
% GENERAL PATH-ENSEMBLE AVERAGES
%%%%%%%%%%%%%%%%%%%%%%%%%%%%%%%%%%%%%%%%%%%%%%%%%%%%%%%%%%%%%%%%%%%%%%

\section{General Path-Ensemble Averages}

Following previous work, \cite{Doss2003,Shirts2008} we estimate nonequilibrium expectations by defining additional path-ensembles with ``unnormalized densities''
\begin{equation}
q_{\mathcal F_i}[X] = \mathcal F[X] \, q_i[X] \:\: ; \:\: c_{\mathcal F_i} = \int dX \, q_{\mathcal F_i}[X] .
\label{eq:functional_density}
\end{equation}
Using Eqs.\ (\ref{eq:path_average}), (\ref{eq:path_density}), and (\ref{eq:functional_density}), we can express nonequilibrium expectations as a ratio of the appropriate normalization constants, $\left< \mathcal F \right>_i = c_{\mathcal F_i} / c_i.$  
Notably, this can be estimated \emph{without} actually sampling path-ensembles biased by some function of $\mathcal F[X]$ (although it is sometimes possible to do so in computer simulations \cite{Sun2003, Ytreberg2004} via transition path sampling \cite{Pratt1986, Dellago1998}).
If no paths are drawn from the path-ensemble corresponding to $q_{\mathcal F_i}[X]$, then $N_{\mathcal F_i} = 0$ and it is no longer required that $q_{\mathcal F_i}[X] > 0$. \cite{Tan2004, Shirts2008}

For each defined path-ensemble, the weight matrix $\bm{M}$ is augmented by one column with elements,
\begin{equation}
M_{n \mathcal F_i} = \hat{c}_{\mathcal F_i}^{-1} \frac{\mathcal F[X_n] \, q_i[X_n] }{ \sum\limits_{k=1}^K \, N_k \, \hat{c}_k^{-1} \, q_k[X_n] } .
\label{eq:F_weight_elements}
\end{equation}
The estimator for the path-ensemble average, $\bar{\mathcal F}_i \approx \left< \mathcal F \right>_i$, can be expressed in terms of weight matrix elements,
\begin{equation}
\bar{\mathcal F}_i = \sum_{n=1}^N \, M_{ni} \, \mathcal F[X_n] ,
\end{equation}
and its uncertainty estimated by
\begin{eqnarray}
\sigma^2 ( \bar{\mathcal F}_i )  &\approx& \bar{\mathcal F}_i^2 (\hat{\Theta}_{\mathcal F_i \, \mathcal F_i} - 2 \hat{\Theta}_{\mathcal F_i \, i} + \hat{\Theta}_{i \, i} ).
\label{eq:path_average_variance}
\end{eqnarray}

%%%%%%%%%%%%%%%%%%%%%%%%%%%%%%%%%%%%%%%%%%%%%%%%%%%%%%%%%%%%%%%%%%%%%%
% Experimentally Relevant Path-Ensembles
%%%%%%%%%%%%%%%%%%%%%%%%%%%%%%%%%%%%%%%%%%%%%%%%%%%%%%%%%%%%%%%%%%%%%%

\section{Experimentally Relevant Path-Ensembles}

The above formalism is fully general, and may be applied to \emph{any} situation where the ratio $q_i[X] / q_j[X]$ can be computed.  For arbitrary path-ensembles, unfortunately, calculating this ratio is only possible in computer simulations unless certain assumptions are made about the dynamics. \cite{Nummela2007}
In a few special path-ensembles, however, we can use the Crooks fluctuation theorem \cite{Crooks1998, Crooks1999} to estimate this ratio, allowing us to apply the EBS estimator to laboratory experiments.  We examine these here.

First, consider a \emph{forward process}, in which a system, initially in equilibrium, is propagated under some time-dependent dynamics for a time $\tau$ which may cause it to be driven out of equilibrium.
The time-dependence of the evolution law (e.g.~Hamiltonian dynamics in a time-dependent potential) is the same for all paths sampled from this ensemble.

For a sample of paths only drawn from this ensemble, the optimal EBS estimator of $\left< \mathcal F \right>_f$ reduces to the sample mean estimator, which we call the \emph{unidirectional} path-ensemble average estimator
\begin{eqnarray}
\bar{\mathcal{F}}_f &=& \frac{1}{N_f} \sum_{n=1}^{N_f} \mathcal F[X_{fn}],
\label{equation:unidirectional-path-estimator}
\end{eqnarray}
and the associated asymptotic variance from Eq.\ \ref{eq:cov} reduces to the variance of the sample mean (see Appendix \ref{sec:uni-var})
\begin{eqnarray}
\sigma^2 ( \bar{\mathcal{F}}_f ) &\approx&
\frac{1}{N_f} \left[ \frac{1}{N_f} \sum_{n=1}^{N_f} \left( \mathcal F[X_{fn}] - \bar{\mathcal F}_f \right)^2 \right]
\end{eqnarray}

The forward process has a unique counterpart known as the \emph{reverse process}.  
Here, the system moves via the opposite protocol in thermodynamic state space; after initial configurations are drawn from the final thermodynamic state of the forward path-ensemble, they are driven towards the initial state.  
If the dynamical law satisfies detailed balance when the control parameters are held constant at each fixed time $t$, the path probabilities in the conjugate forward and reverse path-ensembles are related according to the Crooks fluctuation theorem: \cite{Crooks1998, Crooks1999}
\begin{equation}
\frac{\rho_f[X]}{\rho_r[\tilde{X}]} = \frac{ q_f[X] }{q_r[\tilde{X}] } \frac{ c_r }{ c_f } = e^{w_\tau[X] - \Delta f_\tau} \equiv e^{\Omega[X]},
\label{eq:CFT}
\end{equation}
in which $\tilde{X}$ is the time-reversal, or \emph{conjugate twin}, \cite{Jarzynski2006} of $X$, $\Delta f_t = -\ln (c_t / c_0)$ is the dimensionless free energy difference between thermodynamic states at times $0$ and $t$ (with $\tau$ being the fixed total trajectory length) and $w_t[X]$ is the appropriate dimensionless work.
In Hamiltonian dynamics, for example, this work is $w_t[X] = \beta \int_0^t dt' \, (\partial H/\partial t')$.
For convenience, we define the total \emph{dissipative work} as $\Omega[X] \equiv w_\tau[X] - \Delta f_\tau$.

We will refer to data sets which only include realizations from the forward path-ensemble as `unidirectional', and those with paths from both path-ensembles as `bidirectional'.  Notably, sampling paths from these conjugate ensembles and calculating the associated work $w_t[X]$ is possible in single-molecule pulling experiments as well as computer simulations (c.f.~Refs.~ \cite{Collin2005, Hummer2005}).
To combine bidirectional data to estimate $\left< \mathcal F \right>_f$, we apply the Crooks fluctuation theorem \cite{Crooks1998, Crooks1999} to Eq.\ \ref{eq:opt_bridge} and divide by $\hat{c}_f$, leading to,
\begin{equation}
\bar{\mathcal F}_f = 
\sum_{n=1}^{N_f} \frac{ \mathcal F[X_{fn}] }{ N_f + N_r \, e^{-\hat{\Omega}[X_{fn}] } } + 
\sum_{n=1}^{N_r} \frac{ \mathcal F[X_{rn}] }{ N_f + N_r \, e^{-\hat{\Omega}[X_{rn}] } }
\label{equation:bidirectional-path-estimator}
\end{equation}
which is bidirectional path-average estimator of Minh and Adib, \cite{Minh2008prl} 
derived here by a different route which demonstrates its optimality.
(The asymptotic variance estimator for this equation is written in a closed form in Appendix \ref{sec:bi-var}.)
In these bidirectional expressions, samples drawn from the reverse path-ensemble are time-reversed to obtain the paths $X_{rn}$.  The dissipated work estimate, $\hat{\Omega}[X] \equiv w_\tau[X] - \Delta \hat{f}_\tau$, requires an estimate of $\Delta f_\tau$.  A method for obtaining this estimate will be described next.

%%%%%%%%%%%%%%%%%%%%%%%%%%%%%%%%%%%%%%%%%%%%%%%%%%%%%%%%%%%%%%%%%%%%%%
% ESTIMATES OF EQUILIBRIUM FREE ENERGY
%%%%%%%%%%%%%%%%%%%%%%%%%%%%%%%%%%%%%%%%%%%%%%%%%%%%%%%%%%%%%%%%%%%%%%

\section{Free Energy\label{sec:FE}}

Jarzynski's equality, \cite{Jarzynski1997a, Jarzynski1997b}
\begin{equation}
e^{-\Delta f_t} = \left< e^{-w_t} \right>_f,
\label{eq:JE}
\end{equation}
relates nonequilibrium work and free energy differences.  To facilitate the use of EBS in Jarzynski's equality, we define a path-ensemble by choosing $\mathcal F[X] = e^{-w_t[X]}$ in Eq. \ref{eq:functional_density}, leading to
\begin{equation}
q_{w_t}[X] =  e^{-w_t[X]} \, q_f[X] \:\: ; \:\: c_{w_t} = \int dX \, e^{-w_t[X]} \, q_f[X] .
\end{equation}

When only unidirectional data is available, the optimal EBS estimator for Jarzynski's equality is
\begin{eqnarray}
e^{-\Delta \hat{f}_t} = \frac{1}{N_f} \sum_{n=1}^{N_f} e^{-w_t[X_{fn}]} \label{equation:unidirectional-ft}
\end{eqnarray}
and its asymptotic variance is straightforwardly given by error propagation. \cite{Chipot2007}
Estimators \cite{Sun2003, Ytreberg2004, Minh2009b} and asymptotic variances \cite{Oberhofer2005, Minh2009b} have also been developed for unidirectional \emph{importance sampling} forms of the equality.  

When bidirectional data is available, the same choice of $\mathcal F[X]$ in Eq.\ \ref{equation:bidirectional-path-estimator} gives the estimator
\begin{eqnarray}
e^{-\Delta \hat{f}_t} = 
\sum_{n=1}^{N_f} \frac{ e^{-w_t[X_{fn}]} }{ N_f + N_r \, e^{-\hat{\Omega}[X_{fn}] } } + 
\sum_{n=1}^{N_r} \frac{ e^{-w_t[X_{rn}]} }{ N_f + N_r \, e^{-\hat{\Omega}[X_{rn}] } } \label{equation:bidirectional-ft}
\end{eqnarray}
In this equation, choosing $t=0$ or $t=\tau$ leads to an implicit function mathematically equivalent to the Bennett acceptance ratio method, \cite{Bennett1976, Crooks2000} as previously explained. \cite{Shirts2003, Minh2008prl}
The asymptotic variance of $\Delta \hat{f}_t$ is calculated by augmenting the matrices $\bm{M}$ and $\hat{\bm{\Theta}}$ and using $\phi = \psi = \Delta f_t = - \ln (c_{w_t} / c_f)$ in Eq.\ \ref{eq:cov}, such that,
\begin{equation}
\sigma^2 ( \Delta \hat{f}_t ) = \hat{\Theta}_{w_t \, w_t} - 2\hat{\Theta}_{w_t \, f} + \hat{\Theta}_{f \, f} .
\label{eq:vFt}
\end{equation}

%%%%%%%%%%%%%%%%%%%%%%%%%%%%%%%%%%%%%%%%%%%%%%%%%%%%%%%%%%%%%%%%%%%%%%
% PMF ESTIMATORS FOR SINGLE-MOLECULE PULLING
%%%%%%%%%%%%%%%%%%%%%%%%%%%%%%%%%%%%%%%%%%%%%%%%%%%%%%%%%%%%%%%%%%%%%%

\section{Potential of Mean Force}

Building on Jarzynski's equality, Hummer and Szabo developed expressions for the PMF, \cite{Hummer2001a, Hummer2005} the free energy as a function of a \emph{reaction coordinate} rather than a thermodynamic state, that may be used to interpret single-molecule pulling experiments.  In these experiments, a molecule is mechanically stretched by a force-transducing apparatus, such as an laser optical trap or atomic force microscope tip (c.f. \cite{Hummer2005}).
The Hamiltonian governing the time evolution in these experiments, $H(x;t) = H_0(x) + V(z(x);t)$, is assumed to contain both a term corresponding to the unperturbed system, $H_0(x)$, and a time-dependent (typically harmonic) external bias potential imposed by the apparatus, $V(z;t)$, which acts along a pulling coordinate, $z(x)$.  As the coordinate $z_t \equiv z(x(t))$ is observed at fixed intervals $\Delta t$ over the course of the experiment, we will henceforth use $t = 0,1,...,T$ as an integer time index.  We calculate the work with a discrete sum as $w_t = \sum_{n=1}^{t} \, [V_{n}(z_n) - V_{n-1}(z_n)]$, where $V_n(z) \equiv V(z; n \Delta t)$.

While the expressions in Section \ref{sec:FE} provide an estimate of relative free energies of the equilibrium thermodynamic states defined by $H(x;t)$, they are not immediately useful as an estimate for the PMF along $z$. \cite{Hummer2001a, Hummer2005, Minh2008}  By applying the nonequilibrium estimator for thermodynamic expectations, \cite{Crooks2000, Neal2001} it was shown that the PMF in the absence of an external potential is given by \cite{Hummer2001a, Hummer2005}
\begin{eqnarray}
e^{-g_0(z)} &=& \left< \delta(z - z_t) \, e^{-w_t} \right>_f e^{V(z_t;t)},
\label{eq:HS_time_slice}
\end{eqnarray}
where the dimensionless PMF, $g_0(z)$, is defined in relation to the normalized density as $g_0(z) = - \ln p_0(z) - \delta g$.  In this equation, $\delta g$ is a time-independent constant, $e^{-\delta g} = \int dx ~ e^{-H(x;0)} / \int dx ~  e^{-H_0(x)}$.\cite{Hummer2005}

This theorem can be used to develop estimators for the PMF by replacing the delta function using a kernel function of finite width, such as,
\begin{eqnarray}
h(z-z_t) = 
\begin{cases}
\frac{1}{\Delta z}, & \text{if}~ |z-z_t|< \frac{\Delta z}{2} \\
0, & \text{else}.
\end{cases}
\end{eqnarray}
The width $\Delta z$ must be small so that $e^{V(z;t)}$ does not vary substantially across it.

As this theorem is valid at all times, it is possible to obtain an asymptotically unbiased density estimate $\hat{p}_t$ from each time slice.  It is far more efficient, however, to estimate the PMF using \emph{all} recorded time slices.  While any linear combination of time slices will lead to a valid estimate, certain choices will be more statistically efficient (leading to lower variance) than others.  One way to combine time slices is to use the asymptotic covariance matrix in the method of control variates,\cite{Tan2004} leading to a generalized least-squares optimal estimate of the PMF.  Unfortunately, we empirically found this approach to be numerically unstable.  A more numerically stable approach, which was proposed by Hummer and Szabo, \cite{Hummer2001a, Hummer2005} is based on the weighted histogram analysis method, \cite{Ferrenberg1989,Kumar1992}
\begin{eqnarray}
\hat{p}_0(z) = \frac{\sum_t \mu_t(z) \, \hat{p}_t(z)}{\sum_t \mu_t(z)} \:\: ; \:\: \mu_t(z) \equiv e^{-V(z;t) + \Delta \hat{f}_t } .
\end{eqnarray}
While this weighting scheme is optimal, in a minimal-variance sense, for independent samples from multiple \emph{equilibrium} distributions, these assumptions do not hold for time slices from nonequilibrium trajectories.  However, Oberhofer and Dellago did not observe substantial improvement in PMF estimates when using other time-slice weighting schemes. \cite{Oberhofer2009}

By defining the path-ensemble,
\begin{equation}
q_{z_t}[X] = \delta(z - z_t) \, e^{-w_t[X]} \, q_f[X] \:\: ; \:\: c_{z_t} = \int dX \, q_{z_t}[X]
\end{equation}
and making use of Jarzynski's equality (Eq.\ \ref{eq:JE}) for $e^{-\Delta \hat{f}_t}$, we can write Hummer and Szabo's PMF estimator as
\begin{equation}
e^{-\hat{g}_0(z)} = \frac{ \sum_t (\hat{c}_{z_t} / \hat{c}_{w_t}) }{ \sum_t e^{-V(z;t)} \, (\hat{c}_f / \hat{c}_{w_t}) },
\label{eq:HS}
\end{equation}
which can be readily analyzed in terms of EBS.
While Hummer and Szabo proposed using the unidirectional path average estimator (Eq.\ \ref{equation:unidirectional-path-estimator}) to estimate the expectations in Eq.\ \ref{eq:HS}, Minh and Adib later applied a bidirectional estimator (Eq.\ \ref{equation:bidirectional-path-estimator}), leading to significantly improved statistical properties.  \cite{Minh2008prl}

The asymptotic variance of these estimators can be determined by choosing $\phi = \psi = p_0(z)$ in Eq.\ \ref{eq:cov}.  
For the bidirectional estimator, the matrices $\bm{M}$ and $\hat{\bm{\Theta}}$ will contain one column each for the $f$ and $r$ path-ensembles,  and $T+1$ columns each for the path-ensembles associated with $\{w_t\}_{t=0}^\T$ and $\{z_t\}_{t=0}^\T$.  
The relevant partial derivatives are,
\begin{eqnarray}
\frac{\partial p_0(z)}{\partial \gamma_{f}} & = & 
-p_0(z) \\
\frac{\partial p_0(z)}{\partial \gamma_{w_t}} & = & 
-\frac{1}{\mathcal D} \frac{c_{z_t}}{c_{w_t}} + 
\frac{\mathcal N}{\mathcal D^2} \left( e^{-V(z;t)} \frac{c_f}{c_{w_t}} \right) \\
\frac{\partial p_0(z)}{\partial \gamma_{z_t}} & = & 
\frac{1}{\mathcal D} \frac{c_{z_t}}{c_{w_t}},
\end{eqnarray}
where $\gamma_i = \ln c_i$, $\mathcal N = \sum_t (c_{z_t}/c_{w_t})$ is the numerator of Eq.\ \ref{eq:HS}, and $\mathcal D = \sum_t e^{-V(z;t)} \, (c_f/c_{w_t})$ is its denominator.  
These lead to an estimate for $\sigma^2 (\hat{p}_0(z))$.  
Finally, the asymptotic variance in the PMF is given by the error propagation formula, $\sigma^2 (\hat{g}_0(z)) \approx \sigma^2 (\hat{p}_0(z)) / \hat{p}_0(z)^2$.

%%%%%%%%%%%%%%%%%%%%%%%%%%%%%%%%%%%%%%%%%%%%%%%%%%%%%%%%%%%%%%%%%%%%%%
% ILLUSTRATIVE EXAMPLE
%%%%%%%%%%%%%%%%%%%%%%%%%%%%%%%%%%%%%%%%%%%%%%%%%%%%%%%%%%%%%%%%%%%%%%

\section{Illustrative Example}

We demonstrate these results with Brownian dynamics simulations on a one-dimensional potential with $U_0(z) = (5z^3 - 10z + 3)z$, which were run as previously described. \cite{Minh2008prl}  A time-dependent external perturbation, $V(z;t) = k_s(z-\bar{z}(t))^2/2$, with $k_s = 15$ is applied, such that the total potential is $U(z;t) = U_0(z) + V(z;t)$.  After 100 steps of equilibration at the initial $\bar{z}(t)$, $\bar{z}(t)$ is linearly moved over 750 steps from $-1.5$ to $1.5$ in forward processes and $1.5$ to $-1.5$ in the reverse.  The position at each time step is calculated using the equation $z_t = z_{t-1} - \frac{ dU(x_{t-1}) }{dx} D\Delta t + (2D\Delta t)^{1/2} R_t$, where the diffusion coefficient is $D = 1$, the time step is $\Delta t = 0.001$, and $R_t \sim N(0,1)$ is a random number from the standard normal distribution.

\begin{figure}
\begin{center}
\includegraphics{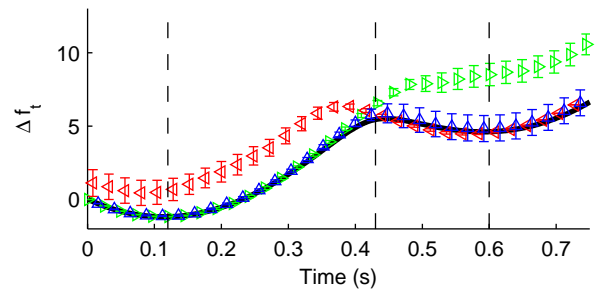}
\caption{\label{fig:Ft}
Comparison of estimators for $\Delta f_t$:  This figure is similar to Fig. 1 of Ref. ~\cite{Minh2008prl}, except that error bars are now included and the sample size is halved.
The unidirectional estimator (Eq.\ \ref{equation:unidirectional-ft}) is applied to 250 forward (rightward triangles) or reverse (leftward triangles, time reversed) sampled paths, and the bidirectional estimator (Eq.\ \ref{equation:bidirectional-ft}) to 125 paths in each direction (upward triangles).  
The exact $\Delta f_t$ is shown as a solid line.  
Error bars (sometimes smaller than the markers) denote one standard deviation of $\Delta \hat f_t$, estimated using the expressions presented here.
The vertical dashed lines are at the times considered in Fig.\ (\ref{fig:error_validation}).}
\end{center}
\end{figure}

As previously noted, \cite{Zuckerman2002, Gore2003, Zuckerman2004, Minh2008prl} unidirectional sampling leads to significant apparent bias in estimates of $\Delta f_t$ (Fig. \ref{fig:Ft}).  
In addition to the increased bias as the system is driven further from equilibrium, we further observe that the estimated variance also increases. 
Bidirectional sampling, on the other hand, leads to a significant reduction in bias and variance, \cite{Minh2008prl} such that free energy estimate is within error bars of the actual free energy.  
Because $\Delta \hat{f}_t$ represents the estimated free energy difference with respect to $t$, the estimated $\sigma^2 ( \Delta \hat{F}_t )$ increases with $t$, becoming equal to the well-known Bennett acceptance ratio asymptotic variance estimate \cite{Bennett1976, Shirts2003} when $t = \tau$.

\begin{figure}
\begin{center}
\includegraphics{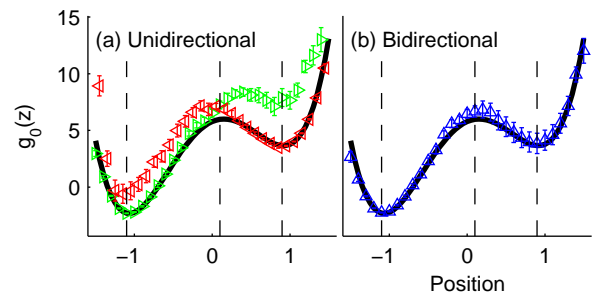}
\caption{\label{fig:pmf}
Comparison of PMF estimators:  This figure is similar to Fig 2 of Ref. ~\cite{Minh2008prl}, except that error bars are now included and the sample size is halved.
In the left panel, the unidirectional Hummer and Szabo estimator is applied to (a) 250 forward (rightward triangles) or 250 reverse (leftward triangles) sampled paths.
In the right panel, the bidirectional estimator is applied to 125 sampled paths in each direction (upward triangles).
The exact PMF is shown as a solid line in both panels.
Error bars (sometimes smaller than the markers) denote one standard deviation of $\Delta \hat g_0(z)$, estimated using the expressions presented here.
The vertical dashed lines are at the positions considered in Fig.\ (\ref{fig:error_validation}).}
\end{center}
\end{figure}

Similar trends are observed with the Hummer-Szabo PMF estimates (Fig. \ref{fig:pmf}).  For unidirectional sampling, the finite-sampling bias and estimated variance increases when the PMF is far from the region sampled by the initial state.  With bidirectional sampling, the bias is significantly reduced; the PMF estimate is largely within error bars of the actual PMF.

\begin{figure}
\begin{center}
\includegraphics{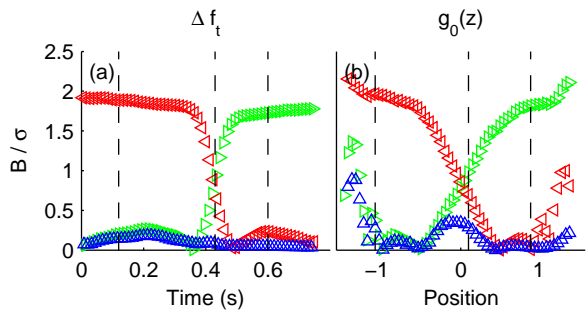}
\caption{\label{fig:bias_vs_variance}
Ratio of estimator bias to standard deviation:  This ratio is calculated for the (a) free energy and (b) PMF, using 1000 independent estimates.  Each estimate is obtained and the type of path sample is indicated as in Figs.\ (\ref{fig:Ft}) and (\ref{fig:pmf}).
The vertical dashed lines are at the times/positions considered in Fig.\ (\ref{fig:error_validation}).
}
\end{center}
\end{figure}

To analyze these trends more quantitatively, we repeated the experiment 1000 times.  For both $\Delta f_t$ and $g_0(z)$, we calculated the bias as 
$\bar{B}(\bar{\mathcal F}_f) = \frac{1}{S} \sum_{s=1}^{S} (\bar{\mathcal F}_{f,s} - \left< \mathcal F \right>)$
and the standard deviation as 
$\bar{\sigma}(\bar{\mathcal F}_f) = \sqrt{ \frac{1}{S} \sum_{s=1}^{S} (\bar{\mathcal F}_{f,s} - \left< \mathcal F \right> )^2 }$, where $S = 1000$ is the number of replicates.  
The results from these more extensive simulations support our described trends (Fig.\ \ref{fig:bias_vs_variance}).  
For unidirectional sampling, the bias in both $\Delta f_t$ and $g_0(x)$ appear to significantly increase around the barrier crossing.  
In the bidirectional free energy estimate, however, the bias is small relative to the variance at all times.  Notably, in the bidirectional PMF estimate, there is a small spike in the bias near the barrier, potentially due to reduced sampling in the region.

While in the large sample limit, the bias in the unidirectional estimate is expected to be small compared to the variance, \cite{Gore2003}
our distribution of unidirectional $e^{-\Delta f_t}$ estimates is significantly skewed and does not resemble a Gaussian distribution expected by the central limit theorem (data not shown).
Hence, the asymptotic limit has not been reached and the large relative bias is caused by insufficient sampling of rare events with low work values that dominate the exponential average. \cite{Jarzynski2006}
Larger sample sizes would be necessary for the distribution of estimates to be normally distributed and for the error to be dominated by the variance (which we estimate here) rather than the bias.

\begin{figure}
\begin{center}
\includegraphics{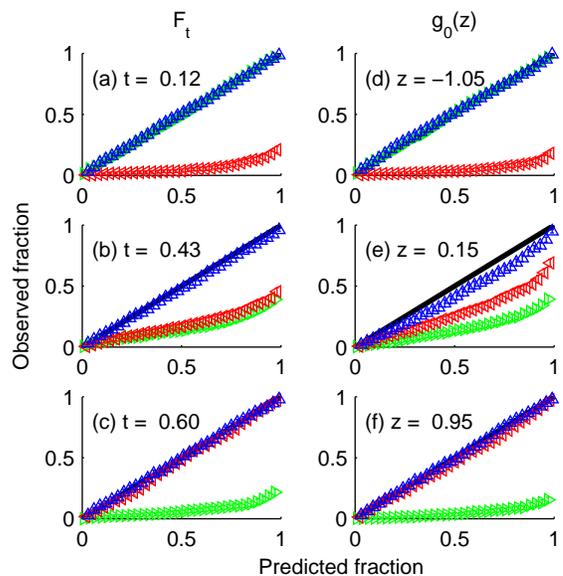}
\caption{\label{fig:error_validation}
Validation of asymptotic variance estimators: Predicted vs. observed fraction of 1000 independent estimates that are within an interval of the true value for (a)-(c) $\Delta f_t$ and (d)-(f) $g_0(z)$ at the indicated times or positions.  Each estimate is obtained and the type of path sample is indicated as in Figs.\ (\ref{fig:Ft}) and (\ref{fig:pmf}).  Error bars on these fractions are 95\% confidence intervals calculated using a Bayesian scheme described in Appendix B of Chodera et. al., ~\cite{Chodera2007} except that, for numerical reasons, the confidence interval was estimated from the variance of the Beta distribution assuming approximate normality, rather than from the inverse Beta cumulative distribution function.
}
\end{center}
\end{figure}

The accuracy of variance estimates may be assessed by comparing predicted and observed confidence intervals.  If the estimates are indeed normally distributed about the true value, about 68\% of estimates from many independent replicates of the experiment should be within one standard deviation of the true value, 95\% within two, and so forth.  Fig.\ (\ref{fig:error_validation}) compares confidence intervals predicted using the described asymptotic variance estimators and the actual fraction of estimates within the interval.

We observe that the accuracy of our asymptotic variance estimate in characterizing the confidence interval largely depends on the presence of bias.  In the bidirectional $\Delta \hat{f}_t$ estimate, where there is little bias, the asymptotic variance estimate works very well.  For the unidirectional $\Delta \hat{f}_t$ estimates, it works well near the initial state but underestimates the error as the system is driven further away from equilibrium, concurring with the bias trend.  In the bidirectional PMF estimate, the asymptotic variance estimate accurately describes the confidence interval except near the barrier, where it slightly underestimates the uncertainty, probably due to the small spike in bias.

In the regime where the bias is much smaller than the variance, $\bar B \ll \bar \sigma$, the asymptotic variance estimate provides a good estimate of the actual statistical error in the estimate.
This also permits us to model the posterior distribution of quantity being estimated as a multivariate normal distribution with mean $\bar{\mathcal F}$ and covariance $\hat{\bm{\Theta}}$.
Doing so provides a route to combining estimates from independent datasets collected from different path ensembles --- such as different pulling speeds or from equilibrium and nonequilibrium path ensembles --- without knowledge of path probability ratios.  This is achieved by maximizing the product of these posterior distributions in a manner similar to the Bayesian approach for estimating $\Delta f_\tau$ described in Ref. ~\cite{Maragakis2008}.

\section{Acknowledgements}
We thank Attila Szabo and Zhiqiang Tan for helpful discussions, and Christopher Calderon for useful comments on the manuscript.  D.M. thanks Artur Adib for supporting a postdoctoral fellowship.  This research was supported by the Intramural Research Program of the NIH, NIDDK.

\begin{widetext}
\begin{appendix}

%%%%%%%%%%%%%%%%%%%%%%%%%%%%%%%%%%%%%%%%%%%%%%%%%%%%%%%%%%%%%%%%%%%%%%
% APPENDIX: CLOSED-FORM ERROR EXPRESSION FOR UNIDIRECTIONAL ESTIMATOR
%%%%%%%%%%%%%%%%%%%%%%%%%%%%%%%%%%%%%%%%%%%%%%%%%%%%%%%%%%%%%%%%%%%%%%

\section{Closed-form expression for the asymptotic variance, given unidirectional data \label{sec:uni-var}}

In this appendix, we show that given unidirectional data, the optimal EBS estimate is the sample mean and its variance simplifies to the variance of a sample mean.  
First, consider the application of the optimal EBS estimator, Eq.\ \ref{eq:opt_bridge}, to estimating a nonequilibrium path-ensemble average from a unidirectional data set,
\begin{eqnarray}
\hat{c}_{\mathcal F_f} & = & 
\sum_{n=1}^{N_f} \left[ \frac{N_f}{\hat{c}_f} \frac{q_f[X_{fn}]}{q_{\mathcal F_f}[X_{fn}]} \right]^{-1} \\
& = & \sum_{n=1}^{N_f} \frac{ \mathcal F[X_{fn}] \hat{c}_f }{N_f}.
\end{eqnarray}
Dividing both sides by $\hat{c}_f$, we obtain the sample mean estimator,
\begin{eqnarray}
\bar{\mathcal F}_f = \frac{1}{N_f} \sum_{n=1}^{N_f} \mathcal F[X_{fn}].
\end{eqnarray}

We shall now simplify the asymptotic variance estimate by closely following the procedure of Shirts and Chodera. \cite{Shirts2008}
When $\bm{M}$ has full column rank, $\hat{\bm{\Theta}}$ can be written as
(Eq.\ D7 of Ref.\ \cite{Shirts2008}),
\begin{eqnarray}
\hat{\bm{\Theta}} = [(\bm{M}^\T\bm{M})^{-1} - \bm{N} + b \bm{1}_K \bm{1}_K^\T]^{-1},
\end{eqnarray}
where $b$ is an arbitrary multiplicative factor and $\bm{1}_K$ is a $1~X~K$ matrix of ones.

The weight matrix $\bm{M}$ consists of two columns,
\begin{eqnarray}
M_{nf} & = & \frac{ \hat{c}_f^{-1} q_f[X_{fn}] }{ N_f \hat{c}_f^{-1} q_f[X_{fn}] } = \frac{1}{N_f} \\
M_{n \mathcal F_f} & = & 
\frac{ \hat{c}_{\mathcal F_f}^{-1} q_{\mathcal F_f}[X_{fn}] }{N_f \hat{c}_f^{-1} q_f[X_{fn}] } =
\frac{\mathcal F[X_{fn}] }{ N_f \bar{\mathcal F}_f },
\end{eqnarray}
obtained by applying Eqs. \ref{eq:weight_elements} and \ref{eq:F_weight_elements}.  
This leads to,
\begin{eqnarray}
\bm{M}^\T\bm{M} = 
\left[
\begin{array}{cc} 
N_f^{-1} & N_f^{-1} \\ 
N_f^{-1} & \sum_{n=1}^{N_f} M_{n \mathcal F_f}^2
\end{array}
\right]
\equiv
\left[
\begin{array}{cc}
a_{11} & a_{12} \\
a_{21} & a_{22}
\end{array}
\right].
\end{eqnarray}
The matrix $\bm{M}^\T\bm{M}$ has the determinant,
\begin{eqnarray}
D = \frac{1}{N_f} \sum_{n=1}^{N_f} M_{n \mathcal F_f}^2 - \frac{1}{N_f^2}.
\end{eqnarray}
which allows us to write the inverse covariance matrix as,
\begin{eqnarray}
\hat{\bm{\Theta}}^{-1} = 
\left[
\begin{array}{cc}
\frac{a_{22}}{D} - N_f + b & -\frac{a_{21}}{D} + b \\
-\frac{a_{12}}{D} + b & \frac{a_{11}}{D} + b
\end{array}
\right].
\end{eqnarray}
By applying the same steps as Appendix E of Shirts and Chodera, \cite{Shirts2008} we then obtain the determinant
\begin{eqnarray}
| \hat{\bm{\Theta}}^{-1} | = \frac{4ab}{D},
\end{eqnarray}
where $a = a_{12} = a_{21}$.  We then obtain the asymptotic covariance estimate,
\begin{eqnarray}
\hat{\bm{\Theta}} = 
\frac{D}{4ab} 
\left[
\begin{array}{cc}
\frac{a_{11}}{D} + b & \frac{a}{D} - b \\
\frac{a}{D} - b & \frac{a_{22}}{D} - N_f + b
\end{array}
\right]
\end{eqnarray}

To estimate the variance, we apply Eq.\ \ref{eq:path_average_variance}, leading to
\begin{eqnarray}
\sigma^2(\bar{\mathcal F}_f^2) 
& \approx & \bar{\mathcal F}_f^2 (\Theta_{\mathcal F_f \, \mathcal F_f} - 2 \Theta_{\mathcal F_f \, f} + \Theta_{f \, f}) \\
& = & \bar{\mathcal F}_f^2 \left(\sum_{n=1}^{N_f} M_{n \mathcal F_f}^2 - \frac{1}{N_f} \right) \\
& = & \sum_{n=1}^{N_f} \frac{ \mathcal F[X_{fn}]^2 }{N_f^2} - \frac{\bar{\mathcal F}_f^2}{N_f}  \\
& = & \frac{1}{N_f} \left[ \frac{1}{N_f} \sum_{n=1}^{N_f}  \mathcal F[X_{fn}]^2 -
\left( \frac{1}{N_f} \sum_{n=1}^{N_f} \mathcal F[X_{fn}] \right)^2 \right]  \\
& = & \frac{1}{N_f} \left[ \frac{1}{N_f} \sum_{n=1}^{N_f} \left( \mathcal F[X_{fn}] - \bar{\mathcal F}_f \right)^2 \right],
\end{eqnarray}
which is the variance of a sample mean estimate.

%%%%%%%%%%%%%%%%%%%%%%%%%%%%%%%%%%%%%%%%%%%%%%%%%%%%%%%%%%%%%%%%%%%%%%
% APPENDIX: CLOSED-FORM ERROR EXPRESSION FOR BIDIRECTIONAL ESTIMATOR
%%%%%%%%%%%%%%%%%%%%%%%%%%%%%%%%%%%%%%%%%%%%%%%%%%%%%%%%%%%%%%%%%%%%%%

\section{Closed-form expression for the asymptotic variance, given bidirectional data \label{sec:bi-var}}

In this appendix, we obtain a closed-form expression for the asymptotic variance of the optimal EBS estimate for $\bar{\mathcal F}_f$, given bidirectional data.  We will follow a similar procedure as in Appendix \ref{sec:uni-var}.  For the bidirectional case, the weight matrix $\bm{M}$ consists of three columns, $\bm{M} = [\bm{m}_f \, \bm{m}_r \, \bm{m}_{\mathcal F_f}]$, where $\bm{m}_i$ is a column matrix of weights from Eqs. \ref{eq:weight_elements} and \ref{eq:F_weight_elements} corresponding to path-ensemble $i$.  The elements of $\bm{M}$ are,
\begin{eqnarray}
M_{nf} & = & \frac{ \hat{c}_f^{-1} q_f[X_n] }{ N_f \hat{c}_f^{-1} q_f[X_n] + N_r \hat{c}_r^{-1} q_r[\tilde{X}_n] } 
= \frac{ 1 }{ N_f  + N_r e^{-\hat{\Omega}[X_n]} } 
= N_f^{-1} \epsilon(L_n) \\
M_{nr} & = & \frac{ \hat{c}_r^{-1} q_r[\tilde{X}_{fn}] }{ N_f \hat{c}_f^{-1} q_f[X_n] + N_r \hat{c}_r^{-1} q_r[\tilde{X}_n] } 
= \frac{ 1 }{ N_f e^{\hat{\Omega}[X_n] } + N_r }
= N_r^{-1} \epsilon(-L_n) \\
M_{n \mathcal F_f} & = & 
\left( \frac{\mathcal F[X] }{\bar{\mathcal F}_f} \right) \frac{ 1}{ N_f  + N_r e^{-\hat{\Omega}[X_n]} }
= \left( \frac{\mathcal F[X] }{\bar{\mathcal F}_f} \right) N_f^{-1} \epsilon(L_n),
\end{eqnarray}
where $\epsilon$ is defined as the Fermi function, $\epsilon(L_n) = \frac{1}{1+e^{-L_n}}$, and we define $L_n = W[X_n] - \Delta \hat{f}_t + \ln \left( \frac{N_f}{N_r} \right)$.  This allows us to write $\bm{M}^\T \bm{M}$ as,
\begin{eqnarray}
\bm{M}^\T\bm{M} & = & 
\sum_{n=1}^N
\left[
\begin{array}{ccc} 
\frac{1}{N_f^2} \epsilon(L_n)^2 & \frac{1}{N_f N_r} \epsilon(L_n) \epsilon(-L_n) 
& \frac{1}{N_f^2} \left( \frac{\mathcal F[X] }{\bar{\mathcal F}_f} \right) \epsilon(L_n)^2 \\
\frac{1}{N_f N_r} \epsilon(L_n) \epsilon(-L_n) & \frac{1}{N_r^2} \epsilon(-L_n)^2
& \frac{1}{N_f N_r} \left( \frac{\mathcal F[X] }{\bar{\mathcal F}_f} \right) \epsilon(L_n) \epsilon(-L_n) \\
 \frac{1}{N_f^2} \left( \frac{\mathcal F[X] }{\bar{\mathcal F}_f} \right) \epsilon(L_n)^2
& \frac{1}{N_f N_r} \left( \frac{\mathcal F[X] }{\bar{\mathcal F}_f} \right) \epsilon(L_n) \epsilon(-L_n)
& \frac{1}{N_f^2} \left( \frac{\mathcal F[X] }{\bar{\mathcal F}_f} \right)^2 \epsilon(L_n)^2
\end{array}
\right] \nonumber \\
& \equiv & 
\left[
\begin{array}{ccc}
a_{ff} & a_{fr} & a_{f \mathcal F_f} \\
a_{fr} & a_{rr} & a_{r \mathcal F_f} \\
a_{\mathcal F_f \mathcal F} & a_{r \mathcal F_f} & a_{\mathcal F_f \mathcal F_f} \\
\end{array}
\right].
\end{eqnarray}
Using the determinant,
\begin{eqnarray}
D = -a_{\mathcal F_f \mathcal F_f} a_{fr}^2 
+ 2 a_{f \mathcal F_f} a_{fr} a_{r \mathcal F_f} 
- a_{ff} a_{r \mathcal F_f}^2 
- a_{f \mathcal F_f}^2 a_{rr} 
+ a_{\mathcal F_f \mathcal F_f} a_{ff} a_{rr},
\end{eqnarray}
we write the inverse covariance matrix estimator as,
\begin{eqnarray}
\hat{\bm{\Theta}}^{-1} =
\left[
\begin{array}{ccc}
\frac{ -a_{r \mathcal F_f}^2 + a_{\mathcal F_f \mathcal F_f} a_{rr} }{D} - N_f + b 
& \frac{ -a_{\mathcal F_f \mathcal F_f} a_{fr} + a_{f \mathcal F_f} a_{r \mathcal F_f} }{D} + b 
& \frac{ a_{fr} a_{r \mathcal F_f} - a_{f \mathcal F_f} a_{rr} }{D}  + b \\
\frac{ -a_{\mathcal F_f \mathcal F_f} a_{fr} + a_{f \mathcal F_f} a_{r \mathcal F_f} }{D} + b 
& \frac{ -a_{f \mathcal F_f}^2 + a_{\mathcal F_f \mathcal F_f} a_{ff} }{D} - N_r + b 
& \frac{ a_{f \mathcal F_f} a_{fr} - a_{ff} a_{r\mathcal F_f} }{D}  + b \\
\frac{ a_{fr} a_{r \mathcal F_f} - a_{f \mathcal F_f} a_{rr} }{D}  + b
& \frac{ a_{f \mathcal F_f} a_{fr} - a_{ff} a_{r\mathcal F_f} }{D}  + b
& \frac{ -a_{fr}^2 + a_{ff} a_{rr} }{D}  + b
\end{array}
\right].
\end{eqnarray}
\end{appendix}

By applying the same steps as Appendix E of Shirts and Chodera, \cite{Shirts2008} we obtain the determinant
\begin{eqnarray}
| \hat{\bm{\Theta}}^{-1} | = \frac{9b(a_{fr}^2 N_f + a_{fr} a_{rr} N_f) }{D}.
\end{eqnarray}

Applying Eq.\ \ref{eq:path_average_variance} to $\hat{\bm{\Theta}}$ and simplifying, it can be shown that the variance estimate is,
\begin{eqnarray}
\sigma^2(\bar{\mathcal F}_f^2) 
& \approx & \bar{\mathcal F}_f^2 (\Theta_{\mathcal F_f \, \mathcal F_f} - 2 \Theta_{\mathcal F_f \, f} + \Theta_{f \, f}) \\
& = & 
\frac{ a_{\mathcal F_f \mathcal F_f} a_{fr} 
- a_{f \mathcal F_f} a_{fr} 
- a_{f \mathcal F_f} a_{r \mathcal F_f}
+ a_{ff} a_{r \mathcal F_f}}
{a_{fr} (a_{fr} N_f +  a_{rr} N_r)}
\end{eqnarray}

\end{widetext}

% \bibliography{asvariance}

\begin{thebibliography}{49}
\expandafter\ifx\csname natexlab\endcsname\relax\def\natexlab#1{#1}\fi
\expandafter\ifx\csname bibnamefont\endcsname\relax
  \def\bibnamefont#1{#1}\fi
\expandafter\ifx\csname bibfnamefont\endcsname\relax
  \def\bibfnamefont#1{#1}\fi
\expandafter\ifx\csname citenamefont\endcsname\relax
  \def\citenamefont#1{#1}\fi
\expandafter\ifx\csname url\endcsname\relax
  \def\url#1{\texttt{#1}}\fi
\expandafter\ifx\csname urlprefix\endcsname\relax\def\urlprefix{URL }\fi
\providecommand{\bibinfo}[2]{#2}
\providecommand{\eprint}[2][]{\url{#2}}

\bibitem[{\citenamefont{Jarzynski}(1997{\natexlab{a}})}]{Jarzynski1997a}
\bibinfo{author}{\bibfnamefont{C.}~\bibnamefont{Jarzynski}},
  \bibinfo{journal}{Phys. Rev. Lett.} \textbf{\bibinfo{volume}{78}},
  \bibinfo{pages}{2690} (\bibinfo{year}{1997}{\natexlab{a}}).

\bibitem[{\citenamefont{Jarzynski}(1997{\natexlab{b}})}]{Jarzynski1997b}
\bibinfo{author}{\bibfnamefont{C.}~\bibnamefont{Jarzynski}},
  \bibinfo{journal}{Phys. Rev. E} \textbf{\bibinfo{volume}{56}},
  \bibinfo{pages}{5018} (\bibinfo{year}{1997}{\natexlab{b}}).

\bibitem[{\citenamefont{Crooks}(2000)}]{Crooks2000}
\bibinfo{author}{\bibfnamefont{G.~E.} \bibnamefont{Crooks}},
  \bibinfo{journal}{Phys. Rev. E} \textbf{\bibinfo{volume}{61}},
  \bibinfo{pages}{2361} (\bibinfo{year}{2000}).

\bibitem[{\citenamefont{Neal}(2001)}]{Neal2001}
\bibinfo{author}{\bibfnamefont{R.~M.} \bibnamefont{Neal}},
  \bibinfo{journal}{Statistics and Computing} \textbf{\bibinfo{volume}{11}},
  \bibinfo{pages}{125} (\bibinfo{year}{2001}).

\bibitem[{\citenamefont{Hummer and Szabo}(2001)}]{Hummer2001a}
\bibinfo{author}{\bibfnamefont{G.}~\bibnamefont{Hummer}} \bibnamefont{and}
  \bibinfo{author}{\bibfnamefont{A.}~\bibnamefont{Szabo}},
  \bibinfo{journal}{Proc. Natl. Acad. Sci. U.S.A.}
  \textbf{\bibinfo{volume}{98}}, \bibinfo{pages}{3658} (\bibinfo{year}{2001}).

\bibitem[{\citenamefont{Hummer and Szabo}(2005)}]{Hummer2005}
\bibinfo{author}{\bibfnamefont{G.}~\bibnamefont{Hummer}} \bibnamefont{and}
  \bibinfo{author}{\bibfnamefont{A.}~\bibnamefont{Szabo}},
  \bibinfo{journal}{Acc. Chem. Res.} \textbf{\bibinfo{volume}{38}},
  \bibinfo{pages}{504} (\bibinfo{year}{2005}).

\bibitem[{\citenamefont{Minh}(2006)}]{Minh2006}
\bibinfo{author}{\bibfnamefont{D.~D.~L.} \bibnamefont{Minh}},
  \bibinfo{journal}{Phys. Rev. E} \textbf{\bibinfo{volume}{74}},
  \bibinfo{pages}{061120} (\bibinfo{year}{2006}).

\bibitem[{\citenamefont{Minh}(2007)}]{Minh2007}
\bibinfo{author}{\bibfnamefont{D.~D.~L.} \bibnamefont{Minh}},
  \bibinfo{journal}{J. Phys. Chem. B} \textbf{\bibinfo{volume}{111}},
  \bibinfo{pages}{4137} (\bibinfo{year}{2007}).

\bibitem[{\citenamefont{Junier et~al.}(2009)\citenamefont{Junier, Mossa,
  Manosas, and Ritort}}]{Junier2009}
\bibinfo{author}{\bibfnamefont{I.}~\bibnamefont{Junier}},
  \bibinfo{author}{\bibfnamefont{A.}~\bibnamefont{Mossa}},
  \bibinfo{author}{\bibfnamefont{M.}~\bibnamefont{Manosas}}, \bibnamefont{and}
  \bibinfo{author}{\bibfnamefont{F.}~\bibnamefont{Ritort}},
  \bibinfo{journal}{Phys. Rev. Lett.} \textbf{\bibinfo{volume}{102}},
  \bibinfo{pages}{070602} (\bibinfo{year}{2009}).

\bibitem[{\citenamefont{Lyman and Zuckerman}(2007)}]{Lyman2007}
\bibinfo{author}{\bibfnamefont{E.}~\bibnamefont{Lyman}} \bibnamefont{and}
  \bibinfo{author}{\bibfnamefont{D.~M.} \bibnamefont{Zuckerman}},
  \bibinfo{journal}{J. Chem. Phys.} \textbf{\bibinfo{volume}{127}},
  \bibinfo{eid}{065101} (pages~\bibinfo{numpages}{6}) (\bibinfo{year}{2007}).

\bibitem[{\citenamefont{Nummela et~al.}(2008)\citenamefont{Nummela, Yassin, and
  Andricioaei}}]{Nummela2008}
\bibinfo{author}{\bibfnamefont{J.}~\bibnamefont{Nummela}},
  \bibinfo{author}{\bibfnamefont{F.}~\bibnamefont{Yassin}}, \bibnamefont{and}
  \bibinfo{author}{\bibfnamefont{I.}~\bibnamefont{Andricioaei}},
  \bibinfo{journal}{J. Chem. Phys.} \textbf{\bibinfo{volume}{128}},
  \bibinfo{eid}{024104} (\bibinfo{year}{2008}).

\bibitem[{\citenamefont{Feng and Crooks}(2009)}]{Feng2009}
\bibinfo{author}{\bibfnamefont{E.~H.} \bibnamefont{Feng}} \bibnamefont{and}
  \bibinfo{author}{\bibfnamefont{G.~E.} \bibnamefont{Crooks}},
  \bibinfo{journal}{Phys. Rev. E} \textbf{\bibinfo{volume}{79}},
  \bibinfo{pages}{012104} (\bibinfo{year}{2009}).

\bibitem[{\citenamefont{Collin et~al.}(2005)\citenamefont{Collin, Ritort,
  Jarzynski, Smith, Tinoco, and Bustamante}}]{Collin2005}
\bibinfo{author}{\bibfnamefont{D.}~\bibnamefont{Collin}},
  \bibinfo{author}{\bibfnamefont{F.}~\bibnamefont{Ritort}},
  \bibinfo{author}{\bibfnamefont{C.}~\bibnamefont{Jarzynski}},
  \bibinfo{author}{\bibfnamefont{S.~B.} \bibnamefont{Smith}},
  \bibinfo{author}{\bibfnamefont{I.}~\bibnamefont{Tinoco}}, \bibnamefont{and}
  \bibinfo{author}{\bibfnamefont{C.}~\bibnamefont{Bustamante}},
  \bibinfo{journal}{Nature} \textbf{\bibinfo{volume}{437}},
  \bibinfo{pages}{231} (\bibinfo{year}{2005}).

\bibitem[{\citenamefont{Liphardt et~al.}(2002)\citenamefont{Liphardt, Dumont,
  Smith, {Tinoco~Jr.}, and Bustamante}}]{Liphardt2002}
\bibinfo{author}{\bibfnamefont{J.}~\bibnamefont{Liphardt}},
  \bibinfo{author}{\bibfnamefont{S.}~\bibnamefont{Dumont}},
  \bibinfo{author}{\bibfnamefont{S.~B.} \bibnamefont{Smith}},
  \bibinfo{author}{\bibfnamefont{I.}~\bibnamefont{{Tinoco~Jr.}}},
  \bibnamefont{and}
  \bibinfo{author}{\bibfnamefont{C.}~\bibnamefont{Bustamante}},
  \bibinfo{journal}{Science} \textbf{\bibinfo{volume}{296}},
  \bibinfo{pages}{1832} (\bibinfo{year}{2002}).

\bibitem[{\citenamefont{Bennett}(1976)}]{Bennett1976}
\bibinfo{author}{\bibfnamefont{C.~H.} \bibnamefont{Bennett}},
  \bibinfo{journal}{J. Comput. Phys.} \textbf{\bibinfo{volume}{22}},
  \bibinfo{pages}{245} (\bibinfo{year}{1976}).

\bibitem[{\citenamefont{Maragakis et~al.}(2006)\citenamefont{Maragakis,
  Spichty, and Karplus}}]{Maragakis2006}
\bibinfo{author}{\bibfnamefont{P.}~\bibnamefont{Maragakis}},
  \bibinfo{author}{\bibfnamefont{M.}~\bibnamefont{Spichty}}, \bibnamefont{and}
  \bibinfo{author}{\bibfnamefont{M.}~\bibnamefont{Karplus}},
  \bibinfo{journal}{Phys. Rev. Lett.} \textbf{\bibinfo{volume}{96}},
  \bibinfo{pages}{100602} (\bibinfo{year}{2006}).

\bibitem[{\citenamefont{Maragakis et~al.}(2008)\citenamefont{Maragakis, Ritort,
  Bustamante, Karplus, and Crooks}}]{Maragakis2008}
\bibinfo{author}{\bibfnamefont{P.}~\bibnamefont{Maragakis}},
  \bibinfo{author}{\bibfnamefont{F.}~\bibnamefont{Ritort}},
  \bibinfo{author}{\bibfnamefont{C.}~\bibnamefont{Bustamante}},
  \bibinfo{author}{\bibfnamefont{M.}~\bibnamefont{Karplus}}, \bibnamefont{and}
  \bibinfo{author}{\bibfnamefont{G.~E.} \bibnamefont{Crooks}},
  \bibinfo{journal}{J. Chem. Phys.} \textbf{\bibinfo{volume}{129}},
  \bibinfo{pages}{024102} (\bibinfo{year}{2008}).

\bibitem[{\citenamefont{Shirts and Pande}(2005)}]{Shirts2005}
\bibinfo{author}{\bibfnamefont{M.~R.} \bibnamefont{Shirts}} \bibnamefont{and}
  \bibinfo{author}{\bibfnamefont{V.~S.} \bibnamefont{Pande}},
  \bibinfo{journal}{J. Chem. Phys.} \textbf{\bibinfo{volume}{122}},
  \bibinfo{pages}{144107} (\bibinfo{year}{2005}).

\bibitem[{\citenamefont{Calderon et~al.}(2009)\citenamefont{Calderon, Janosi,
  and Kosztin}}]{Calderon2009}
\bibinfo{author}{\bibfnamefont{C.~P.} \bibnamefont{Calderon}},
  \bibinfo{author}{\bibfnamefont{L.}~\bibnamefont{Janosi}}, \bibnamefont{and}
  \bibinfo{author}{\bibfnamefont{I.}~\bibnamefont{Kosztin}},
  \bibinfo{journal}{J. Chem. Phys.} \textbf{\bibinfo{volume}{130}},
  \bibinfo{eid}{144908} (pages~\bibinfo{numpages}{13}) (\bibinfo{year}{2009}).

\bibitem[{\citenamefont{Tan}(2004)}]{Tan2004}
\bibinfo{author}{\bibfnamefont{Z.}~\bibnamefont{Tan}}, \bibinfo{journal}{J. Am.
  Stat. Assoc.} \textbf{\bibinfo{volume}{99}}, \bibinfo{pages}{1027}
  (\bibinfo{year}{2004}).

\bibitem[{\citenamefont{Singhal and Pande}(2005)}]{Singhal2005}
\bibinfo{author}{\bibfnamefont{N.}~\bibnamefont{Singhal}} \bibnamefont{and}
  \bibinfo{author}{\bibfnamefont{V.~S.} \bibnamefont{Pande}},
  \bibinfo{journal}{J. Chem. Phys.} \textbf{\bibinfo{volume}{123}},
  \bibinfo{pages}{204909} (\bibinfo{year}{2005}).

\bibitem[{\citenamefont{Hinrichs and Pande}(2007)}]{Singhal2007}
\bibinfo{author}{\bibfnamefont{N.~S.} \bibnamefont{Hinrichs}} \bibnamefont{and}
  \bibinfo{author}{\bibfnamefont{V.~S.} \bibnamefont{Pande}},
  \bibinfo{journal}{J. Chem. Phys.} \textbf{\bibinfo{volume}{126}},
  \bibinfo{pages}{244101} (\bibinfo{year}{2007}).

\bibitem[{\citenamefont{Hahn and Then}(2009)}]{Hahn2009}
\bibinfo{author}{\bibfnamefont{A.~M.} \bibnamefont{Hahn}} \bibnamefont{and}
  \bibinfo{author}{\bibfnamefont{H.}~\bibnamefont{Then}},
  \emph{\bibinfo{title}{A dynamic sampling strategy for two-sided free-energy
  estimation}} (\bibinfo{year}{2009}), \eprint{cond-mat/0904.0625v2}.

\bibitem[{\citenamefont{Vardi}(1985)}]{Vardi1985}
\bibinfo{author}{\bibfnamefont{Y.}~\bibnamefont{Vardi}}, \bibinfo{journal}{Ann.
  Stat.} \textbf{\bibinfo{volume}{13}}, \bibinfo{pages}{178}
  (\bibinfo{year}{1985}).

\bibitem[{\citenamefont{Gill et~al.}(1988)\citenamefont{Gill, Vardi, and
  Wellner}}]{Gill1988}
\bibinfo{author}{\bibfnamefont{R.~D.} \bibnamefont{Gill}},
  \bibinfo{author}{\bibfnamefont{Y.}~\bibnamefont{Vardi}}, \bibnamefont{and}
  \bibinfo{author}{\bibfnamefont{J.~A.} \bibnamefont{Wellner}},
  \bibinfo{journal}{Ann. Stat.} \textbf{\bibinfo{volume}{16}},
  \bibinfo{pages}{1069} (\bibinfo{year}{1988}).

\bibitem[{\citenamefont{Kong et~al.}(2003)\citenamefont{Kong, McCullagh, Meng,
  Nicolae, and Tan}}]{Kong2003}
\bibinfo{author}{\bibfnamefont{A.}~\bibnamefont{Kong}},
  \bibinfo{author}{\bibfnamefont{P.}~\bibnamefont{McCullagh}},
  \bibinfo{author}{\bibfnamefont{X.-L.} \bibnamefont{Meng}},
  \bibinfo{author}{\bibfnamefont{D.}~\bibnamefont{Nicolae}}, \bibnamefont{and}
  \bibinfo{author}{\bibfnamefont{Z.}~\bibnamefont{Tan}}, \bibinfo{journal}{J.
  R. Stat. Soc. Ser. B (Stat. Methodol.)} \textbf{\bibinfo{volume}{65}},
  \bibinfo{pages}{585} (\bibinfo{year}{2003}).

\bibitem[{\citenamefont{Shirts and Chodera}(2008)}]{Shirts2008}
\bibinfo{author}{\bibfnamefont{M.~R.} \bibnamefont{Shirts}} \bibnamefont{and}
  \bibinfo{author}{\bibfnamefont{J.~D.} \bibnamefont{Chodera}},
  \bibinfo{journal}{J. Chem. Phys.} \textbf{\bibinfo{volume}{129}},
  \bibinfo{pages}{124105} (\bibinfo{year}{2008}).

\bibitem[{\citenamefont{Minh and Adib}(2008)}]{Minh2008prl}
\bibinfo{author}{\bibfnamefont{D.~D.~L.} \bibnamefont{Minh}} \bibnamefont{and}
  \bibinfo{author}{\bibfnamefont{A.~B.} \bibnamefont{Adib}},
  \bibinfo{journal}{Phys. Rev. Lett.} \textbf{\bibinfo{volume}{100}},
  \bibinfo{pages}{180602} (\bibinfo{year}{2008}).

\bibitem[{Dos()}]{Doss2003}
\bibinfo{note}{Honi Doss makes this suggestion in the conference discussion
  of~\protect\cite{Kong2003}}.

\bibitem[{\citenamefont{Sun}(2003)}]{Sun2003}
\bibinfo{author}{\bibfnamefont{S.}~\bibnamefont{Sun}}, \bibinfo{journal}{J.
  Chem. Phys.} \textbf{\bibinfo{volume}{118}}, \bibinfo{pages}{5769}
  (\bibinfo{year}{2003}).

\bibitem[{\citenamefont{Ytreberg and Zuckerman}(2004)}]{Ytreberg2004}
\bibinfo{author}{\bibfnamefont{F.~M.} \bibnamefont{Ytreberg}} \bibnamefont{and}
  \bibinfo{author}{\bibfnamefont{D.~M.} \bibnamefont{Zuckerman}},
  \bibinfo{journal}{J. Chem. Phys.} \textbf{\bibinfo{volume}{120}},
  \bibinfo{pages}{10876} (\bibinfo{year}{2004}).

\bibitem[{\citenamefont{Pratt}(1986)}]{Pratt1986}
\bibinfo{author}{\bibfnamefont{L.}~\bibnamefont{Pratt}}, \bibinfo{journal}{J.
  Chem. Phys.} \textbf{\bibinfo{volume}{85}}, \bibinfo{pages}{5045}
  (\bibinfo{year}{1986}).

\bibitem[{\citenamefont{Dellago et~al.}(1998)\citenamefont{Dellago, Bolhuis,
  Csajka, and Chandler}}]{Dellago1998}
\bibinfo{author}{\bibfnamefont{C.}~\bibnamefont{Dellago}},
  \bibinfo{author}{\bibfnamefont{P.~G.} \bibnamefont{Bolhuis}},
  \bibinfo{author}{\bibfnamefont{F.~S.} \bibnamefont{Csajka}},
  \bibnamefont{and} \bibinfo{author}{\bibfnamefont{D.}~\bibnamefont{Chandler}},
  \bibinfo{journal}{J. Chem. Phys.} \textbf{\bibinfo{volume}{108}},
  \bibinfo{pages}{1964} (\bibinfo{year}{1998}).

\bibitem[{\citenamefont{Nummela and Andricioaei}(2007)}]{Nummela2007}
\bibinfo{author}{\bibfnamefont{J.}~\bibnamefont{Nummela}} \bibnamefont{and}
  \bibinfo{author}{\bibfnamefont{I.}~\bibnamefont{Andricioaei}},
  \bibinfo{journal}{Biophys. J.} \textbf{\bibinfo{volume}{93}},
  \bibinfo{pages}{3373} (\bibinfo{year}{2007}).

\bibitem[{\citenamefont{Crooks}(1998)}]{Crooks1998}
\bibinfo{author}{\bibfnamefont{G.~E.} \bibnamefont{Crooks}},
  \bibinfo{journal}{J. Stat. Phys.} \textbf{\bibinfo{volume}{90}},
  \bibinfo{pages}{1481} (\bibinfo{year}{1998}).

\bibitem[{\citenamefont{Crooks}(1999)}]{Crooks1999}
\bibinfo{author}{\bibfnamefont{G.~E.} \bibnamefont{Crooks}},
  \bibinfo{journal}{Phys. Rev. E} \textbf{\bibinfo{volume}{60}},
  \bibinfo{pages}{2721} (\bibinfo{year}{1999}).

\bibitem[{\citenamefont{Jarzynski}(2006)}]{Jarzynski2006}
\bibinfo{author}{\bibfnamefont{C.}~\bibnamefont{Jarzynski}},
  \bibinfo{journal}{Phys. Rev. E} \textbf{\bibinfo{volume}{73}},
  \bibinfo{pages}{046105} (\bibinfo{year}{2006}).

\bibitem[{\citenamefont{Lu and Woolf}(2007)}]{Chipot2007}
\bibinfo{author}{\bibfnamefont{L.}~\bibnamefont{Lu}} \bibnamefont{and}
  \bibinfo{author}{\bibfnamefont{T.~B.} \bibnamefont{Woolf}}, in
  \emph{\bibinfo{booktitle}{Free Energy Calculations}}, edited by
  \bibinfo{editor}{\bibfnamefont{C.}~\bibnamefont{Chipot}} \bibnamefont{and}
  \bibinfo{editor}{\bibfnamefont{A.}~\bibnamefont{Pohorille}}
  (\bibinfo{publisher}{Springer, Berlin}, \bibinfo{year}{2007}),
  vol.~\bibinfo{volume}{86}.

\bibitem[{\citenamefont{Minh}(2009)}]{Minh2009b}
\bibinfo{author}{\bibfnamefont{D.~D.~L.} \bibnamefont{Minh}},
  \bibinfo{journal}{J. Chem. Phys.} \textbf{\bibinfo{volume}{130}},
  \bibinfo{pages}{204102} (\bibinfo{year}{2009}).

\bibitem[{\citenamefont{Oberhofer et~al.}(2005)\citenamefont{Oberhofer,
  Dellago, and Geissler}}]{Oberhofer2005}
\bibinfo{author}{\bibfnamefont{H.}~\bibnamefont{Oberhofer}},
  \bibinfo{author}{\bibfnamefont{C.}~\bibnamefont{Dellago}}, \bibnamefont{and}
  \bibinfo{author}{\bibfnamefont{P.}~\bibnamefont{Geissler}},
  \bibinfo{journal}{J. Phys. Chem. B} \textbf{\bibinfo{volume}{109}},
  \bibinfo{pages}{6902} (\bibinfo{year}{2005}).

\bibitem[{\citenamefont{Shirts et~al.}(2003)\citenamefont{Shirts, Bair, Hooker,
  and Pande}}]{Shirts2003}
\bibinfo{author}{\bibfnamefont{M.~R.} \bibnamefont{Shirts}},
  \bibinfo{author}{\bibfnamefont{E.}~\bibnamefont{Bair}},
  \bibinfo{author}{\bibfnamefont{G.}~\bibnamefont{Hooker}}, \bibnamefont{and}
  \bibinfo{author}{\bibfnamefont{V.~S.} \bibnamefont{Pande}},
  \bibinfo{journal}{Phys. Rev. Lett.} \textbf{\bibinfo{volume}{91}},
  \bibinfo{pages}{140601} (\bibinfo{year}{2003}).

\bibitem[{\citenamefont{Minh and McCammon}(2008)}]{Minh2008}
\bibinfo{author}{\bibfnamefont{D.~D.~L.} \bibnamefont{Minh}} \bibnamefont{and}
  \bibinfo{author}{\bibfnamefont{J.~A.} \bibnamefont{McCammon}},
  \bibinfo{journal}{J. Phys. Chem. B} \textbf{\bibinfo{volume}{112}},
  \bibinfo{pages}{5892} (\bibinfo{year}{2008}).

\bibitem[{\citenamefont{Ferrenberg and Swendsen}(1989)}]{Ferrenberg1989}
\bibinfo{author}{\bibfnamefont{A.~M.} \bibnamefont{Ferrenberg}}
  \bibnamefont{and} \bibinfo{author}{\bibfnamefont{R.~H.}
  \bibnamefont{Swendsen}}, \bibinfo{journal}{Phys. Rev. Lett.}
  \textbf{\bibinfo{volume}{63}}, \bibinfo{pages}{1195} (\bibinfo{year}{1989}).

\bibitem[{\citenamefont{Kumar et~al.}(1992)\citenamefont{Kumar, Bouzida,
  Swendsen, Kollman, and Rosenberg}}]{Kumar1992}
\bibinfo{author}{\bibfnamefont{S.}~\bibnamefont{Kumar}},
  \bibinfo{author}{\bibfnamefont{D.}~\bibnamefont{Bouzida}},
  \bibinfo{author}{\bibfnamefont{R.~H.} \bibnamefont{Swendsen}},
  \bibinfo{author}{\bibfnamefont{P.~A.} \bibnamefont{Kollman}},
  \bibnamefont{and} \bibinfo{author}{\bibfnamefont{J.~M.}
  \bibnamefont{Rosenberg}}, \bibinfo{journal}{J. Comput. Chem.}
  \textbf{\bibinfo{volume}{13}}, \bibinfo{pages}{1011} (\bibinfo{year}{1992}).

\bibitem[{\citenamefont{Oberhofer and Dellago}(2009)}]{Oberhofer2009}
\bibinfo{author}{\bibfnamefont{H.}~\bibnamefont{Oberhofer}} \bibnamefont{and}
  \bibinfo{author}{\bibfnamefont{C.}~\bibnamefont{Dellago}},
  \bibinfo{journal}{J. Comput. Chem.} \textbf{\bibinfo{volume}{30}},
  \bibinfo{pages}{1726} (\bibinfo{year}{2009}).

\bibitem[{\citenamefont{Zuckerman and Woolf}(2002)}]{Zuckerman2002}
\bibinfo{author}{\bibfnamefont{D.~M.} \bibnamefont{Zuckerman}}
  \bibnamefont{and} \bibinfo{author}{\bibfnamefont{T.~B.} \bibnamefont{Woolf}},
  \bibinfo{journal}{Phys. Rev. Lett.} \textbf{\bibinfo{volume}{89}},
  \bibinfo{pages}{180602} (\bibinfo{year}{2002}).

\bibitem[{\citenamefont{Gore et~al.}(2003)\citenamefont{Gore, Ritort, and
  Bustamante}}]{Gore2003}
\bibinfo{author}{\bibfnamefont{J.}~\bibnamefont{Gore}},
  \bibinfo{author}{\bibfnamefont{F.}~\bibnamefont{Ritort}}, \bibnamefont{and}
  \bibinfo{author}{\bibfnamefont{C.}~\bibnamefont{Bustamante}},
  \bibinfo{journal}{Proc. Natl. Acad. Sci. U.S.A.}
  \textbf{\bibinfo{volume}{100}}, \bibinfo{pages}{12564}
  (\bibinfo{year}{2003}).

\bibitem[{\citenamefont{Zuckerman and Woolf}(2004)}]{Zuckerman2004}
\bibinfo{author}{\bibfnamefont{D.~M.} \bibnamefont{Zuckerman}}
  \bibnamefont{and} \bibinfo{author}{\bibfnamefont{T.~B.} \bibnamefont{Woolf}},
  \bibinfo{journal}{J. Stat. Phys.} \textbf{\bibinfo{volume}{114}},
  \bibinfo{pages}{1303} (\bibinfo{year}{2004}).

\bibitem[{\citenamefont{Chodera et~al.}(2007)\citenamefont{Chodera, Swope,
  Pitera, Seok, and Dill}}]{Chodera2007}
\bibinfo{author}{\bibfnamefont{J.~D.} \bibnamefont{Chodera}},
  \bibinfo{author}{\bibfnamefont{W.~C.} \bibnamefont{Swope}},
  \bibinfo{author}{\bibfnamefont{J.~W.} \bibnamefont{Pitera}},
  \bibinfo{author}{\bibfnamefont{C.}~\bibnamefont{Seok}}, \bibnamefont{and}
  \bibinfo{author}{\bibfnamefont{K.~A.} \bibnamefont{Dill}},
  \bibinfo{journal}{J. Chem. Theory Comput.} \textbf{\bibinfo{volume}{3}},
  \bibinfo{pages}{26} (\bibinfo{year}{2007}).

\end{thebibliography}

\end{document}